\documentclass[12pt,a4paper]{article}
\usepackage{latexsym,amssymb,amsmath,amsthm}
\usepackage[english]{babel}

\newtheorem{theorem}{Theorem}
\newtheorem{lemma}{Lemma}

\newcommand{\OR}{\mathsf{OR}}

\begin{document}
\title{On relative OR-complexity of Boolean matrices and their complements
\footnote{Research is supported in part by RFBR, grant
14--01--00671a.}}
\author{Igor S. Sergeev\footnote{e-mail: isserg@gmail.com}}
\date{}
\maketitle

We construct explicit Boolean square matrices whose rectifier
complexity (OR-complexity) differs significantly from the
complexity of the complement matrices. This note can be viewed as
an addition to the material of~\cite[\S 5.6]{js}.

Recall that {\it rectifier $(m,n)$-circuit} is an oriented graph
with $n$ vertices labeled as inputs and $m$ vertices labeled as
outputs. {\it Rectifier circuit} (OR-{\it{circuit}}) implements a
Boolean $m\times n$ matrix $A=(A[i,j])$ iff for any $i$ and $j$
the value $A[i,j]$ indicates the existence of an oriented path
from $j$-th input to $i$-th output. Complexity of a circuit is the
number of edges in it, circuit depth is the maximal length of an
oriented path. See details in~\cite{js,lu}.

We denote by $\OR(A)$ the complexity of an edge-minimal circuit
imple\-men\-ting a given matrix $A$; if we speak about circuits of
depth $\le d$, then the corresponding complexity is denoted by
$\OR_d(A)$.

It was proved in~\cite{js} via method~\cite{k} the existence of
$n\times n$-matrices $A$ satisfying
$$ \OR(\bar A)/\OR(A) = \Omega(n/\log^3 n). $$
Note that due to general results~\cite{lu,n} on the asymptotic
complexity of the class of Boolean matrices the ratio in the
question cannot exceed $\Theta(n/\log n)$.

A {\it $k$-rectangle} is an all-ones $k\times k$ matrix. A matrix
is {\it $k$-free} if it does not contain a $k$-rectangle as a
submatrix.

It was established in~\cite{js} the existence of an $n\times n$
matrix $A$ simple for depth-2 circuits, $\OR_2(A) = O(n\log^2 n)$,
whose complement matrix $\bar A$ is 2-free and has relatively high
weight (the number of ones) $|\bar A| = \Omega(n^{5/4})$. As a
consequence of~\cite{n}, $ \OR(\bar A) = \OR_2(\bar A) = |\bar
A|.$

Below, we provide an explicit construction of matrices satisfying
similar conditions.

\begin{theorem} $(i)$ For an explicit Boolean $n \times n$ matrix~$C$:
$$ \OR(\bar C)/\OR(C) = n \cdot 2^{-O(\sqrt{\ln n \ln\ln n})}.$$

$(ii)$ For an explicit Boolean $n \times n$ matrix~$C$ the
following conditions hold: $\OR(C) = O(n)$, matrix $\bar C$ is
$2$-free and $|\bar C| = \Omega(n^{4/3})$.
\end{theorem}

(Recall that the weight of any 2-free matrix is at most
$n^{3/2}+n$.)

The proof of the theorem is based on the following simple
combinatorial lemma.

\begin{lemma}
Let the weight of an $n\times n$ matrix $A$ be $|A| \ge 2n^{3/2}$.
Then $A$ contains $\Omega((|A|/n)^4)$ $2$-rectangles.
\end{lemma}

\proof Say that a row {\it covers} a pair $u$ of two columns, if
this row has ones in these columns. If $a_i$ denotes the number of
ones in the $i$-th row of $A$, then the number of pairs of columns
covered by the rows of $A$ is
$$
\sigma = \sum_{i=1}^n \binom{a_i}{2} = \frac12\sum_{i=1}^n a_i^2
-\frac{|A|}2 \ge \frac{\left(\sum_{i=1}^n
a_i\right)^2}{2n}-\frac{|A|}2 = \frac{|A|^2}{2n}-\frac{|A|}2 \ge
\frac{|A|^2}{4n}.
$$
Let $b_u$ be the number of rows covering the pair $u$ of columns.
Then $\sum_u b_u = \sigma$. Thus, the number of 2-rectangles in
$A$ is
\begin{multline*}
\sum_u \binom{b_u}2 = \frac12\sum_u b_u^2 -\frac{\sigma}2 \ge
\frac{\left(\sum_u b_u\right)^2}{n(n-1)}-\frac{\sigma}2 = \\ =
\frac{\sigma^2}{n(n-1)}-\frac{\sigma}2 \ge \frac{\sigma^2}{2n^2} =
\Omega\left(\left(\frac{|A|}{n}\right)^4\right).
\end{multline*}
\qed

Let $n=\binom{m}2$. Given an $m\times m$ matrix $A$ construct an
$n\times n$ matrix $B$ as follows. Label rows and columns of $B$
by 2-element subsets of $[m]$. Set $B[a,b]=1$ iff $a \times b$
forms a 2-rectangle in $B$.

\begin{lemma}
If $A$ is $k$-free, then $B$ is $K$-free, $K= \binom{k-1}2+1$.
\end{lemma}

\proof Suppose that $B$ contains a $K$-rectangle at the
intersection of rows $s_1,\ldots,s_K$ and columns
$t_1,\ldots,t_K$. Then $A$ contains a rectangle at the
intersection of rows $\cup s_i$ and columns $\cup t_i$. But
necessarily $|\cup s_i|, |\cup t_i| \ge k$, contradicting
$k$-freeness of $A$. \qed

\begin{lemma}
If $A$ is $k$-free and $|A| \ge 2m^{3/2}$, then
$$ \OR(B) = \Omega\left( \left(\frac{|A|}{kn}\right)^4 \right), $$
on the other hand, $\OR_3(\bar B) = O(n)$.
\end{lemma}

\proof By Lemma 1, $|B| = \Omega((|A|/n)^4)$, and Lemma 2 implies
that $B$ is $K$-free. Therefore, by the Nechiporuk's
theorem~\cite{n}
$$ \OR(B) \ge \frac{|B|}{K^2} = \Omega\left( \left(\frac{|A|}{kn}\right)^4 \right). $$

We are left to show that the matrix $\bar B$ can be implemented by
a depth-3 circuit of linear complexity. Take a depth-3 circuit
where the nodes on the second and the third layer are numbers
$1,\ldots,m$, and there is an edge joining an input or an output
$a$ with a node $i$ iff $i \in a$. The edges between the second
and the third layers are drown according to the entries of the
matrix $\bar A$.

By the construction, the circuit has $O(m^2)$ edges. Indeed, it
implements the matrix~$\bar B$ since there exists a path
connecting an input $a$ with an output $b$ iff the submatrix at
the intersection of rows $b$ and columns $a$ is not all-zero. \qed

To prove p. (i) of the Theorem take $m\times m$ norm-matrix
$A$~\cite{krs}, which is $\Delta$-free and has $m^2/\Delta$ ones,
where $\Delta=2^{O(\sqrt{\log m\log\log m})}$, under appropriate
choice of parameters. Put $C=\bar B$.

To prove p. (ii) take 3-free $m\times m$ Brown's matrix
$A$~\cite{b} of weight $\Theta(m^{5/3})$. Put $C=\bar B$. \qed

The author is grateful to Stasys Jukna for suggestions improving
the presentation.

\end{document}